\documentclass[final,5p,times,twocolumn]{elsarticle}
\usepackage{amsmath,amssymb,amsfonts}
\usepackage{graphicx}
\usepackage{slashed}

\allowdisplaybreaks[1]

\newcommand{\beq}{\begin{equation}}
\newcommand{\eeq}{\end{equation}}
\newcommand{\Order}{\mathcal{O}}
\newcommand{\Lagr}{\mathcal{L}}
\newcommand{\mpp}{m_{\rm p}}
\newcommand{\mn}{m_{\rm n}}
\newcommand{\mpi}{M_{\pi}}
\newcommand{\mpii}{M_{\pi^0}}
\newcommand{\muu}{m_{\rm u}}
\newcommand{\md}{m_{\rm d}}
\newcommand{\Fpi}{F_\pi}
\newcommand{\ga}{g_{\rm A}}

\begin{document}
%%%%%%%%%%%%%%%%%%%%%%%%%%%%%%%%%%%%%%%%%%%%%%%%%%%%%%%%%%%%%%%%%%%%%%%%%%%%%

\renewcommand{\theequation}{\arabic{equation}}

\begin{frontmatter}

\title{\hfill{\scriptsize HISKP--TH--09/11, FZJ-IKP-TH-2009-9}\\ 
Isospin breaking in the pion--nucleon scattering lengths}

\author[Bonn]{Martin~Hoferichter}
\author[Bonn]{Bastian~Kubis}
\author[Bonn,Julich]{Ulf-G.~Mei{\ss}ner}

\address[Bonn]{Helmholtz--Institut f\"ur Strahlen- und Kernphysik (Theorie) 
   and Bethe Center for Theoretical Physics, Universit\"at Bonn, D-53115 Bonn, Germany}
\address[Julich]{Institut f\"ur Kernphysik (Theorie), Institute for Advanced Simulations, 
   and J\"ulich Center for Hadron Physics, Forschungszentrum J\"ulich, D-52425  J\"ulich, Germany}

\begin{abstract}
We analyze isospin breaking through quark mass differences and virtual photons 
in the pion--nucleon scattering lengths in all physical channels in the
framework of covariant baryon chiral perturbation theory. 
\end{abstract}


\end{frontmatter}

\section{Introduction}

Isospin violation in the Standard Model is driven by strong and 
electromagnetic interactions, that is by the differences in the
light quark masses and charges, respectively. As already stressed
by Weinberg, the pion--nucleon scattering lengths offer a particularly
good testing ground for strong isospin violation~\cite{Weinberg77}. This problem
was addressed in the framework of heavy-baryon chiral perturbation
theory (ChPT) in a series of papers about a decade ago 
\cite{MS97,FMS99,MM99,FM01b,FM01}.
Recently, new interest arose in high-precision calculations of the pion--nucleon 
scattering lengths. First, the accurate measurements of the characteristics
of pionic hydrogen and deuterium allow one in principle to extract
certain $\pi N$ scattering lengths to high precision.
This, however, is only possible if isospin breaking is 
taken into account consistently. In the case of the strong energy shift of 
the ground state of pionic hydrogen one needs the isospin-violating 
contributions to $a_{\pi^- p\to \pi^- p}$. In~\cite{GR02}, these have been determined
at third order in the chiral expansion, $\Order\big(p^3\big)$, 
in a covariantly regularized form of baryon ChPT~\cite{BL99}.
[For a recent review on baryon ChPT, see \cite{Bernard:2007zu}.]
As for the width of pionic hydrogen, the knowledge of the isospin-breaking 
corrections to $a_{\pi^- p\to \pi^0 n}$ is required. 
In the analysis of pionic deuterium isospin violation is particularly
important, since the $\pi d$ scattering length at leading order is
proportional to the small isoscalar scattering length $a^+$ 
and therefore chirally suppressed (cf.~\cite{MRR06}). Since
$\text{Re}\,a_{\pi d}\propto a_{\pi^- p\to \pi^- p}+ a_{\pi^- n\to \pi^- n}
+\text{few-body corrections},$ we may improve at least the two-body 
contributions by extending the isospin-breaking corrections to 
$a_{\pi^-n\rightarrow \pi^- n}$ to $\Order(p^3)$. Second, as has been 
stressed in particular by Bernstein, threshold pion photoproduction offers
the unique possibility of measuring the so far undetermined $\pi^0 p$ scattering
length and gives access to the charge exchange scattering length $a_{\pi^+ n
\to\pi^0 p}$, see~\cite{Bernstein:1998ip} and the recent
review~\cite{Bernstein:2009dc}. Such measurements are becoming 
feasible at HI$\gamma$S and at MAMI.
In view of these developments, it is timely to
extend the work of~\cite{GR02} to {\it all} charge channels in
pion--nucleon scattering. 

\section{Formalism}

We start the description of various formal aspects of $\pi N$ scattering at threshold with the kinematics.
The momenta of the nucleon and pion in the initial (final) state will be denoted by $p$ ($p'$) and $q$ ($q'$), respectively, 
their masses by $m_{\rm i}$ ($m_{\rm f}$) and $M_{\rm i}$ ($M_{\rm f}$). 
$\mpp$, $\mn$, $\mpi$, and $\mpii$, are the masses of proton, neutron, and charged and neutral pion.
We define the isospin limit by the charged particle masses $\mpp$ and $\mpi$.
Working at first order in isospin breaking, i.e.\ at $\Order(e^2,\md-\muu)\equiv\Order(\delta)$,
we only need contributions linear in $\Delta_\pi=\mpi^2-\mpii^2$ and $\Delta_{\rm N}=\mn-\mpp$.

For elastic scattering, the kinematics at threshold are determined by
\beq
 s=(m_{\rm i}+M_{\rm i})^2, \quad p=p'=\frac{m_{\rm i}}{M_{\rm i}}q=\frac{m_{\rm i}}{M_{\rm i}}q', \quad t=0. \label{23}
\eeq
\eqref{23} is modified for the charge exchange reactions (cex) according to
\beq
p \neq p',\quad q \neq q',\quad  t=-\Delta_\pi+\frac{\mpi}{m_{\rm i}+\mpi}(m_{\rm f}^2-m_{\rm i}^2+\Delta_\pi).
\eeq
In loop contributions that only start at $\Order(p^3)$, these kinematical relations may be chirally expanded, leading to
\beq
s=(\mpp+\mpi)^2, \quad t=-\Delta_\pi, \quad p=p'=\frac{\mpp}{\mpi}q.\label{22}
\eeq
Note that still $q'$ must not be replaced by $q$, since the difference is of the same chiral order as $q$ and $q'$ themselves.

The pion--nucleon scattering amplitude $T_{\pi N}$ is parameterized in terms of the two amplitudes
$D(s,t)$ and $B(s,t)$ according to 
\begin{align}
T_{\pi N}&=\bar{u}(p')\bigg(D(s,t)-\frac{1}{2(m_{\rm i}+m_{\rm f})}[\slashed{q}',\slashed{q}]B(s,t)\bigg)u(p),\notag\\
 \bar{u}(p')&u(p')=2 m_{\rm f},\quad \bar{u}(p)u(p)=2m_{\rm i}.
\end{align}
In the isospin limit,  $T_{\pi N}$ may be decomposed as
\beq
T^{ab}=T^{+}\delta^{ab}+T^{-}\frac{1}{2}[\tau^a,\tau^b],
\eeq
where $a$ ($b$) is the isospin index of the outgoing (incoming) pion and $\tau^i$ are the Pauli-matrices. 
Using the Condon--Shortley phase convention, the physical amplitudes are related to $T^+$ and $T^-$ by
\begin{align}
T_{\pi^- p}&\equiv T_{\pi^- p\rightarrow \pi^- p}=T_{\pi^+ n}\equiv T_{\pi^+ n\rightarrow \pi^+ n}=T^++T^-,\notag\\
T_{\pi^+ p}&\equiv T_{\pi^+ p\rightarrow \pi^+ p}=T_{\pi^- n}\equiv T_{\pi^- n\rightarrow \pi^- n}=T^+-T^-,\notag\\
T_{\pi^- p}^{\rm cex}&\equiv T_{\pi^- p\rightarrow \pi^0 n}=T_{\pi^+ n}^{\rm cex}\equiv T_{\pi^+ n\rightarrow \pi^0 p}=-\sqrt{2}\,T^-,\notag\\
T_{\pi^0 p}&\equiv T_{\pi^0 p\rightarrow \pi^0 p}=T_{\pi^0 n}\equiv T_{\pi^0 n\rightarrow \pi^0 n}=T^+.
\end{align}
For the elastic channels only $D(s,t)$ contributes at threshold, whereas we find for the charge exchange reactions
\beq
T_{\pi N}=2\sqrt{\mn\mpp}\left(\left(1+\frac{\Delta_\pi}{8\mpp^2}\right)D_{\rm thr}-\frac{\mpi\Delta_\pi}{4\mpp^2}B_{\rm thr}\right),
\eeq
where $D_{\rm thr}$ and $B_{\rm thr}$ denote the amplitudes evaluated at threshold. 
The correction factor in front of $D$ stems from the expansion of the Dirac spinors around the isospin limit. 
Since the prefactor is already of first order in $\delta$, $B_{\rm thr}$ may be evaluated assuming isospin symmetry 
to relate it to isovector threshold parameters \cite{BL01},
\begin{align}
 B^-_{\rm thr}&=8\pi\mpp\left(\frac{a^-_{0+}}{4\mpp^2}+a^-_{1-}-a^-_{1+}\right)\notag\\
&=\frac{1}{2\Fpi^2}\big(1+4 \mpp c_4\big)+\Order(p),\quad B_{\rm thr}=-\sqrt{2}B^-_{\rm thr},\label{eq:Bthr}
\end{align}
where $a^-_{l\pm}$ denotes the isovector scattering lengths with orbital momentum $l$ 
and total angular momentum $l\pm\frac{1}{2}$. 
For brevity, we will use $a^\pm\equiv a^\pm_{0+}$ for the S-wave isoscalar and isovector scattering lengths. 
Equation~\eqref{eq:Bthr} also shows the leading chiral representation of $B^-_{\rm thr}$.
All relevant terms of the effective chiral Lagrangians defining the corresponding low-energy constants
are collected in Appendix~\ref{app:lagr}.

The S-wave scattering length $a$ for elastic scattering of scalar particles is related to the amplitude $T(s,t)$ by
\beq
a = \frac{1}{8\pi\sqrt{s}}T(s,t)\bigg|_{|\mathbf{p}|\rightarrow 0}\label{19},
\eeq
where $|\mathbf{p}|$ is the center-of-mass momentum. 
This result is generalized to pion--nucleon scattering by
\begin{align}
a_{\rm elastic}&=\frac{m_{\rm i}}{4\pi(m_{\rm i}+M_{\rm i})}D_{\rm thr}^{\rm elastic},\notag\\
a_{\rm cex}&=\frac{\sqrt{\mpp\mn}}{4\pi(m_{\rm i}+\mpi)}\Biggl\{\biggl(1+\frac{\Delta_\pi}{8\mpp^2}\biggr)D^{\rm cex}_{\rm thr}-\frac{\mpi\Delta_\pi}{4\mpp^2}B^{\rm cex}_{\rm thr}\Biggl\}.\label{6}
\end{align}
The isospin-symmetric contributions to the scattering lengths have already been worked out in \cite{BKM93}. 
Adapted to our notation they read
\begin{align}
a^+&=\frac{\mpp\mpi^2}{4\pi(\mpp+\mpi)\Fpi^2}\bigg\{-\frac{\ga^2}{4\mpp}
+2(c_2+c_3-2c_1)+
\frac{3\ga^2\mpi}{64\pi\Fpi^2}\bigg\},\notag\\
a^-&=\frac{\mpp\mpi}{8\pi(\mpp+\mpi)\Fpi^2}\bigg\{1+\frac{\ga^2\mpi^2}{4\mpp^2}+\frac{\mpi^2}{8\pi^2\Fpi^2}\left(1-\log\frac{\mpi^2}{\mu^2}\right)\notag\\
&+8\mpi^2\big(d^{\rm r}_1+d^{\rm r}_2+d^{\rm r}_3+2d^{\rm r}_5\big)+\frac{2\mpi^2}{\Fpi^2}l^{\rm r}_4\bigg\}.\label{1}
\end{align}

As soon as we take into account virtual photons, 
we have to specify more carefully what we mean by $D_{\rm thr}$
due to the appearance of threshold divergences.
First of all, we subtract all one-photon-reducible diagrams, since they diverge $\sim 1/t$, 
and denote the result by $\tilde{D}$. 
The additional divergences due to photon loops may be regularized in the form
\beq
\left.\left(e^{iQ\alpha\theta_{\rm C}(|\mathbf{p}|)}\tilde{D}(s,t)\right)\right|_{|\mathbf{p}|\rightarrow 0}=\frac{\beta_1}{|\mathbf{p}|}+\beta_2\log\frac{|\mathbf{p}|}{\mu_{\rm c}}+D_{\rm thr}+\Order(|\mathbf{p}|)\label{5},
\eeq
where $\alpha={e^2}/{4\pi}$ denotes the fine structure constant,
$\theta_{\rm C}(|\mathbf{p}|)$ the infrared divergent Coulomb phase given by
\beq
\theta_{\rm C}(|\mathbf{p}|)=-\frac{\mu_{\rm c}}{|\mathbf{p}|}\log\frac{m_\gamma}{2|\mathbf{p}|} \,,
\eeq
and $\mu_{\rm c}={\mpp\mpi}/{(\mpp+\mpi)}$ the reduced mass of the incoming particles. 
$Q$ accounts for the charges of the particles involved, explicitly
\beq
Q_{\pi^- p}=-2,\quad Q_{\pi^+ p}=2, \quad Q_{\pi^- p}^{\rm cex}=-1,
\eeq
and $Q=0$ for the remaining channels. 
For consistency reasons, the contribution from $B(s,t)$ to the charge exchange reaction should be multiplied 
by the same phase as $\tilde{D}(s,t)$.
Since $\alpha\theta_{\rm C}(|\mathbf{p}|)$ is of first order in isospin breaking, 
this does not lead to additional terms at the order considered here, 
but makes it obvious that $\theta_{\rm C}(|\mathbf{p}|)$ drops out of physical observables. 
The coefficients $\beta_i$ may be related to the scattering lengths $a$. The explicit relation 
\beq
\beta_1=-\frac{\pi}{2}Q e^2\mpi a
\eeq
is confirmed by our calculation at leading order in the chiral expansion, but can be proven to hold in general 
in the framework of non-relativistic effective field theories \cite{GLR01,BFGKB09}. 
The coefficient $\beta_2$ only appears at two-loop level.

\section{Analytic results}\label{sec:analytic}

\begin{figure}
\begin{center}
\includegraphics[width=\linewidth]{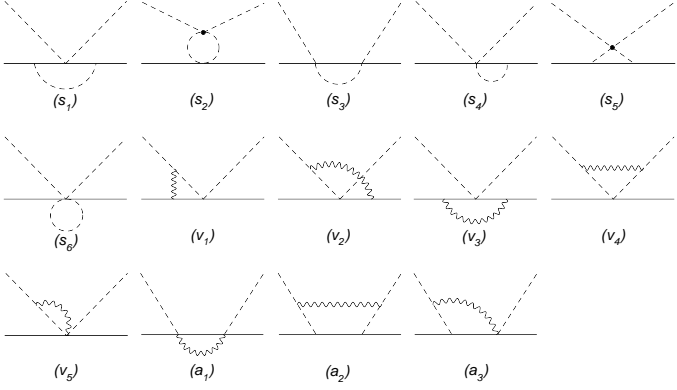}
\end{center}
\caption{Loop diagrams for $\pi N$ scattering at threshold. 
Solid, dashed, and wiggly lines, denote nucleons, pions, and photons, respectively.
Crossed diagrams and diagrams contributing via wave function renormalization only are not shown.}
\label{fig:Feynman}
\end{figure}

The topologies of the Feynman diagrams contributing at threshold are displayed in Fig.~\ref{fig:Feynman}. 
There are significantly less diagrams than above threshold, since many diagrams which are formally of 
$\Order(p^3)$ start only at $\Order(p^4)$ for the following reasons.
Firstly, the leading term of a particular diagram can be determined by simplifying the numerators 
according to chiral power counting, making use of \eqref{23} and \eqref{22}. 
With $\Sigma=p+q$ and loop momentum $k$ a typical example for such a simplification is
\begin{align}
& (\slashed{\Sigma}-\slashed{k}+\mpp)\slashed{q}\gamma_5(\slashed{p}-\slashed{k}+\mpp)\rightarrow  (\slashed{p}+\mpp)\frac{\mpi}{\mpp}\slashed{p}\gamma_5(\slashed{p}+\mpp)\notag\\
&=-\gamma_5\frac{\mpi}{\mpp}\slashed{p}(-\slashed{p}+\mpp)(\slashed{p}+\mpp)=0.
\end{align}
Secondly, all $s$-channel one-particle-reducible diagrams involve structures of the type 
$(\slashed{\Sigma}+\mpp)\slashed{q}\gamma_5u(p)$, whose leading part vanishes at threshold, since
\begin{align}
&(\slashed{\Sigma}+\mpp)\slashed{q}\gamma_5u(p)\rightarrow (\slashed{p}+\mpp)\frac{\mpi}{\mpp}\slashed{p}\gamma_5u(p)\notag\\
&=\frac{\mpi}{\mpp}\slashed{p}\gamma_5(-\slashed{p}+\mpp)u(p)=0.
\end{align}
The $u$-channel diagrams are treated analogously. 
Unfortunately, both arguments only work for $q$ and not for $q'$ in the charge exchange reactions, 
unless the diagram in question is already of order $\Order(\delta)$;
but eventually one can show that all diagrams which may be omitted in the case of the elastic channels 
do not contribute to the charge exchange reactions either.

Concentrating on the analysis of the isospin-breaking shifts in the scattering lengths,
we obtain the following results for the reaction channels on the proton
(the neutron channels can be found in Appendix~\ref{app:chann}):
\begin{align}
\Delta a_{\pi^- p} &= a_{\pi^- p} - (a^+ + a^- )= \Delta a^+ + \Delta a^- + i\, \text{Im}\, a_{\pi^- p} \,,\notag\\
\Delta a_{\pi^+ p} &= a_{\pi^+ p} - (a^+ - a^- )= \Delta a^+ - \Delta a^-  \,,\notag\\
\Delta a^+ &= \frac{\mpp}{4\pi(\mpp+\mpi)}
\Bigg\{\!\frac{4\Delta_\pi}{\Fpi^2}c_1-\frac{e^2}{2}\big(4f_1+f_2\big) \notag\\
&-\frac{\ga^2\mpi}{32\pi\Fpi^2}\biggl(\frac{33\Delta_\pi}{4\Fpi^2} +e^2\biggr)\!\Bigg\} \notag\\
\Delta a^- &=-\frac{\mpp\mpi}{4\pi(\mpp+\mpi)}
\Bigg\{
\frac{\Delta_\pi}{32\pi^2\Fpi^4}\left(3+\log\frac{\mpi^2}{\mu^2}\right)
\notag\\
&+\frac{8\Delta_\pi}{\Fpi^2}d^{\rm r}_5+\frac{e^2\ga^2}{16\pi^2\Fpi^2}\left(1+4\log 2+3\log\frac{\mpi^2}{\mu^2}\right) \notag\\
&-2e^2\left(g^{\rm r}_6+g^{\rm r}_8-\frac{5}{9\Fpi^2}\big(k^{\rm r}_1+k^{\rm r}_2\big)\right)\Bigg\},\notag\\
\text{Im}\, a_{\pi^- p} &=\frac{\mpp}{4\pi(\mpp+\mpi)} \Bigg\{
\frac{\mpi^2}{8\pi\Fpi^4}\sqrt{\Delta_\pi-2\mpi\Delta_{\rm N}}+\frac{e^2\ga^2\mpi}{4\pi\Fpi^2}\Bigg\},\notag\\
 \Delta a_{\pi^- p}^{\rm cex}&=a_{\pi^- p}^{\rm cex}+\sqrt{2}\,a^-=\frac{\sqrt{2}\,\mpp}{4\pi(\mpp+\mpi)}
\Bigg\{\frac{e^2f_2}{2}\notag\\
&+\frac{\ga^2\Delta_\pi}{4\Fpi^2\mpp}+\frac{\mpi\Delta_\pi}{4\mpp^2}\biggl(B^-_{\rm thr}-\frac{3}{4\Fpi^2}\biggr)+\frac{8\mpi\Delta_\pi}{\Fpi^2}d^{\rm r}_5\notag\\
&+\frac{\mpi\Delta_\pi}{192\pi^2\Fpi^4}\left(2-7\ga^2+\big(2-5\ga^2\big)\log\frac{\mpi^2}{\mu^2}\right)\notag\\
&+\frac{e^2\mpi}{32\pi^2\Fpi^2}\left(5+3\log\frac{\mpi^2}{\mu^2}\right)-\frac{\mpi\Delta_{\rm N}}{4\Fpi^2\mpp}\big(1+2\ga^2\big)\notag\\
&+\frac{e^2\mpi}{2\Fpi^2}\left(\Fpi^2 g^{\rm r}_7-2k^{\rm r}_3+k^{\rm r}_4+\frac{20}{9}\big(k^{\rm r}_1+k^{\rm r}_2\big)\right)\Bigg\},\notag\\
 \Delta a_{\pi^0 p}&=a_{\pi^0 p}-a^+=-\frac{\Delta_\pi}{\mpi^2}a^++\frac{\mpp}{4\pi(\mpp+\mpi)}
\Bigg\{\frac{3\ga^2\mpi\Delta_\pi}{128\pi\Fpi^4}\notag\\
&-\frac{\mpi^2\sqrt{\Delta_\pi+2\mpi\Delta_{\rm N}}}{8\pi\Fpi^4}+\frac{2c_5 B(\md-\muu)}{\Fpi^2}\Bigg\}.\label{eq:Delta_a}
\end{align}
We wish to point explicitly to the square-root-like terms in $\text{Im}\, a_{\pi^- p}$ and $\Delta a_{\pi^0 p}$, 
which are caused by the unitarity cusps due to the different thresholds of the $\pi^0n$ and $\pi^+n$ intermediate
states, respectively.  These cusps can be calculated exactly at threshold, 
which we will illustrate for the imaginary part in Sect.~\ref{sec:Im}. 
Since the cusp is of order $\Order\big(\sqrt{\delta}\big)$ and thus enhanced 
compared to the other isospin-breaking effects,
we also take into account the correction by $\Delta_{\rm N}$, although this is formally an $\Order(p^4)$ effect. 
Nevertheless, it contributes $\sim 30\; \%$ to the difference between $a_{\pi^0 p}$ and $a_{\pi^0 n}$
(see Appendix~\ref{app:chann}).

We have performed the following checks on our calculation: the amplitudes are ultraviolet-finite, 
all ultraviolet divergences due to loops are 
canceled by the infinite parts of the counterterms (as calculated in~\cite{GR02}). 
Thus, only the renormalized counterterms appear in \eqref{eq:Delta_a}. 
They compensate the scale dependence generated by the chiral logarithms, such that the final results 
are independent of the renormalization scale $\mu$. 
Furthermore, the infrared divergences caused by virtual photons cancel among themselves, as they should. 

A useful way to quantify isospin-breaking corrections in terms of measurable quantities is
the so-called triangle relation that vanishes in the isospin limit. It is defined as
\beq
R=2\frac{ a_{\pi^+ p}- a_{\pi^- p}-\sqrt{2} a_{\pi^- p}^{\rm cex}}{ a_{\pi^+ p}- a_{\pi^- p}+\sqrt{2} a_{\pi^- p}^{\rm cex}},
\eeq
where only the real parts of the scattering lengths are inserted. At first order in $\delta$ we obtain
\begin{align}
R&=\frac{\mpp}{4\pi (\mpp+\mpi)a^-}\Bigg\{\frac{e^2 f_2}{2}+\frac{\ga^2\Delta_\pi}{4\Fpi^2\mpp}-\frac{\mpi\Delta_{\rm N}}{4\Fpi^2\mpp}\big(1+2\ga^2\big)\notag\\
&-\frac{3\mpi\Delta_\pi}{16\Fpi^2\mpp^2}+\frac{\mpi\Delta_\pi}{4\mpp^2}B^-_{\rm thr}
-\frac{\mpi\Delta_\pi}{48\pi^2\Fpi^4}\left(4+\log\frac{\mpi^2}{\mu^2}\right)\notag\\
&-\frac{\ga^2\mpi\Delta_\pi}{192\pi^2\Fpi^4}\left(7+5\log\frac{\mpi^2}{\mu^2}\right)+\frac{e^2\mpi}{32\pi^2\Fpi^2}\left(5+3\log\frac{\mpi^2}{\mu^2}\right)\notag\\
&-\frac{e^2\ga^2\mpi}{16\pi^2\Fpi^2}\left(1+4\log 2+3\log\frac{\mpi^2}{\mu^2}\right)\notag\\
&+\frac{e^2\mpi}{2}\big(4g_6^{\rm r}+g_7^{\rm r}+4g_8^{\rm r}\big)+\frac{e^2\mpi}{2\Fpi^2}\big(k_4^{\rm r}-2k_3^{\rm r}\big)\Bigg \}.\label{9}
\end{align}
We refrain from constructing an isoscalar triangle relation from the three elastic
pion--proton scattering lengths (cf.\ $R_1$ in~\cite{FMS99}); such a relation can easily
be read off from the results in~\eqref{eq:Delta_a}.  It depends on the low-energy constants
$f_1$, $f_2$, and $c_1$, and, as we will see in the following section, therefore cannot be
very well constrained, such that no additional information
beyond the shifts in the individual scattering lengths is provided.

\section{Numerical results}
\subsection{Low-energy constants}
The most precise values for $a^+$ and $a^-$ 
stem from an analysis of pionic hydrogen and pionic deuterium data~\cite{MRR06}
\beq
a^+=(1.5\pm 2.2)\cdot10^{-3}\mpi^{-1},\quad a^-=(85.2\pm 1.8)\cdot10^{-3}\mpi^{-1} \label{10}.
\eeq
In addition, the authors extract the electromagnetic low-energy constant (LEC)
$
 f_1=-2.1^{+3.2}_{-2.2} \,{\rm GeV}^{-1} .
$
$f_2$ and $c_5$ can be deduced from the mass difference between proton and neutron. 
This mass difference comprises electromagnetic as well as strong contributions 
\beq
 \mn-\mpp=-4Bc_5(\md-\muu)+f_2e^2\Fpi^2 \label{eq:masses},
 \eeq
which may be disentangled by means of the Cottingham formula~\cite{GL82}. The result of this procedure is
$
f_2 =-(0.97\pm0.38)\,{\rm GeV}^{-1}
$,
$
Bc_5 (\md-\muu) =-(0.51\pm0.08)\,{\rm MeV}
$.

In~\cite{M05}, various previous analyses of $c_1$ are briefly reviewed and combined to
$
c_1=- 0.9^{+0.2}_{-0.5} \,{\rm GeV}^{-1}.
$
For $d_5^r$, we will use
$
\Fpi^2 d_5^{\rm r}(\mu)=(0.6\pm 3.0)\cdot 10^{-3},
$
specifying the renormalization scale to $\mu=1\,{\rm GeV}$.
The central value is the mean of the values quoted in~\cite{BL01} (translated to our conventions regarding $\Lagr^{(p^4)}_\pi$),
where a low-energy theorem linking $d_5^r$ to a certain subthreshold parameter of $\pi N$ scattering is derived.
In the spirit of the treatment of $c_1$, we estimate the error by investigating the effects of higher orders 
in this low-energy theorem.
Neglecting the fourth order contribution would shift $\Fpi^2d_5^{\rm r}$ by
\beq
\frac{\mpi}{16}\frac{64 \mpp c_1+\ga^2\big[2(4+\ga^2)+\sqrt{2}\,\log(1+\sqrt{2})\big]}{32\pi\,\mpp} 
=-3\cdot 10^{-3}
\eeq
(for the central value of $c_1$).
The resulting uncertainty ensures consistency with most values for $d_5^{\rm r}$ available in the 
literature~\cite{MOJ98,BM99,FMS98,F00,FM00,FM01}.

We now turn to the determination of $B^-_{\rm thr}$. 
Values for $a^-_{1-}$ and $a^-_{1+}$ can be found in \cite{Ko80,FMS98,F00,FM00}. Using
\beq
a^-_{1-}=(-12\pm 2)\cdot 10^{-3}\mpi^{-3}, \quad a^-_{1+}=(-81\pm 6)\cdot 10^{-3}\mpi^{-3}
\eeq
yields
\beq
B^-_{\rm thr}=(0.60\pm 0.06)\cdot 10^{-3}\,{\rm MeV}^{-2}.
\eeq
Since the main source for the determination of $c_4$ are $\pi N$ threshold data, 
it seems more reliable to apply the threshold parameters directly.

Estimates of the meson-sector electromagnetic LECs $k_i$ are given in \cite{HIS07} 
using resonance saturation \cite{MOU97,AMOU04}. 
Unfortunately, this method does not provide reliable error estimates. The central values for the $k_i$ in question are
$k_1^{\rm r}=10.9\cdot 10^{-3}$, 
$k_2^{\rm r}= 0.7\cdot 10^{-3}$,
$k_3^{\rm r}= 3.9\cdot 10^{-3}$, 
$k_4^{\rm r}=-1.3\cdot 10^{-3}$ (all at $\mu=1\,$GeV).
Since $g_6^{\rm r}$, $g_7^{\rm r}$, and $g_8^{\rm r}$ are not known, they are set to zero in the numerical work. 
Particle masses and decay constants are taken from \cite{PDG08}, in particular $\Fpi=92.2 \,{\rm MeV}$ and $|\ga|=1.2695$.

\subsection{Triangle relation, scattering lengths}
The triangle relation
$R$ can be determined rather well since $f_1$, the $\Order(p^2)$ LEC which is least known, drops out. 
The central value is obtained by inserting the above LECs and \eqref{10} into \eqref{9}. As for the error, 
we are faced with the following combination of electromagnetic LECs whose uncertainty is not known:
\beq
k_4^{\rm r}-2k_3^{\rm r}+\Fpi^2(4g_6^{\rm r}+g_7^{\rm r}+4g_8^{\rm r}).
\eeq
Naively one would assign the order-of-magnitude errors ${1}/{16\pi^2}$
to each LEC and add the individual contributions in quadrature. 
However, this may underestimate the uncertainty in case the variation
of the renormalization scale $\mu$ by a factor of $e=2.718\dots$, controlled by the corresponding $\beta$-functions,
 induces a change significantly larger than ${1}/{16\pi^2}$.
Assuming $1 \,\rm{GeV}$ to be a ``natural'' scale for hadronic processes, 
this running covers the energy range the physics we consider should be sensitive to. 
Estimating the uncertainty by varying the LECs according to their $\beta$-functions
in a correlated way has the further advantage of being independent under 
redefinition of the Lagrangian. The result of this procedure is
\begin{align}
R&=(1.5\pm 0.2_{f_2} \pm 0.03_{a^-} \pm 0.03_{B^-_{\rm thr}} \pm 1.1_{\rm LEC})\,\%\notag\\
&=(1.5\pm 1.1)\,\% \label{16},
\end{align}
where the different contributions to the error are denoted by a subscript, 
``${\rm LEC}$'' standing here and in the following for the corresponding combination of LECs with unknown error. 
The final uncertainty is obtained by adding the individual contributions in quadrature.
Naive order-of-magnitude arguments would reduce the error significantly to $0.4\,\%$.
The large error in \eqref{16} is dominated by the $g_i^{\rm r}$, as may be seen from their $\beta$-functions~\cite{GR02}
$
\eta_7=-{9}/{2}-{2}Z(5\ga^2+1)/3 =-9.4
$,
$
\eta_8=-2\eta_6=3(4\ga^2-1)/2+{2}Z(5\ga^2+1)/3=13.1
$,
which are by no means of order $\Order(1)$. 

We now turn to the isospin-violating contributions to the individual scattering lengths.
The procedure as described above yields 
\begin{align}
 \Delta a_{\pi^- p}
 &\!=\Big(\!-3.4^{+1.2}_{-2.9\,c_1}{}^{+3.9}_{-5.7 \, f_1}\! \pm \! 0.2 _{f_2}\! \pm \!0.6 _{d_5} \!\pm\! 1.2_{\rm LEC}+5.0 i\Big) 10^{-3}\!\mpi^{-1} \notag\\
&\!=\Big(\!-3.4^{+4.3}_{-6.5}+5.0 i\Big)\cdot 10^{-3}\mpi^{-1},\notag\\
\Delta a_{\pi^+ p}
&\!=\Big(\!-5.3 ^{+1.2}_{-2.9\,c_1}{}^{+3.9}_{-5.7 \, f_1} \pm 0.2 _{f_2} \pm 0.6 _{d_5} \pm 1.2_{\rm LEC}\Big)\cdot 10^{-3}\mpi^{-1}\notag\\
&\!=-5.3^{+4.3}_{-6.5}\cdot 10^{-3}\mpi^{-1},\notag\\
\Delta a_{\pi^- p}^{\rm cex}
&\!=\Big(0.4 \pm 0.2_{f_2} \pm 0.8_{d_5} \pm 0.04_{B^-_{\rm thr}} \pm 0.4_{\rm LEC}\Big)\cdot 10^{-3}\mpi^{-1}\notag\\
&\!=\big(0.4\pm0.9\big)\cdot 10^{-3}\mpi^{-1},\notag\\
\Delta a_{\pi^0 p}&\!=\big(\!-5.2 \pm 0.1_{a^+} \pm 0.2_{c_5}\big)\cdot 10^{-3}\mpi^{-1}\notag\\
&\!=\big(\!-5.2 \pm 0.2\big)\cdot 10^{-3}\mpi^{-1} .\label{12}
\end{align}
Discarding the imaginary part, \eqref{12} corresponds to relative changes compared to the isospin limit of 
$-3.9^{+4.9}_{-7.5}\,\%$ in $a_{\pi^- p}$,
$+6.4^{+7.8}_{-5.1}\,\%$ in $a_{\pi^+ p}$, and
$(-0.4 \pm 0.8)\,\%$ in $a_{\pi^- p}^{\rm cex}$.
Due to the poor knowledge of $a^+$, the corresponding normalization of $\Delta a_{\pi^0 p}$ is not very meaningful;
note that the isospin-breaking shift $\Delta a_{\pi^0 p}$ in \eqref{12} is significantly
larger than $a^+$.
As already pointed out in~\cite{GR02}, the large isospin-breaking corrections
to the charged-pion elastic channels can be traced back to the triangle graph $(s_{5})$ (see Fig.~\ref{fig:Feynman}),
which however only yields a rather small contribution to the charge exchange reaction.
In contrast, isospin violation in the neutral-pion elastic channel is predominantly due to the cusp effect enhanced by $\sqrt{\delta}$.
The large uncertainties in $\Delta a_{\pi^\pm p}$ are dominated by $f_1$ and $c_1$ 
that are part of $\Delta a^+$ in \eqref{eq:Delta_a}, therefore appear in the same combination 
with $a^+$ in both channels.

\subsection{Imaginary parts}\label{sec:Im}

Exact expressions for the imaginary parts of $a_{\pi^- p}$ and $a_{\pi^- p}^{\rm cex}$ generated by the 
$\pi^0 n$ and $\gamma n$ intermediate states can be obtained using Cutkosky rules, 
expressing the vertices at threshold by scattering lengths and electric dipole amplitudes 
$E_{0+}$ encountered in the context of pion photoproduction. 
Retaining all chiral orders, the resulting imaginary parts up to $\Order(\delta)$ are
\begin{align}
\text{Im}\begin{Bmatrix}a_{\pi^- p}\\a_{\pi^- p}^{\rm cex}\end{Bmatrix}
&= \frac{a^-\sqrt{2\mpp}}{\sqrt{\mpp+\mpi}}\sqrt{\Delta_\pi-2\mpi\Delta_{\rm N}}
\begin{Bmatrix}\sqrt{2}\,a^-\\-a^+\end{Bmatrix}\notag\\
&+\frac{\mpi E_{0+}(\pi^-p)}{\big(\mpp+\mpi\big)}\big(\mpi+2\mpp\big)
\begin{Bmatrix}E_{0+}(\pi^-p)\\E_{0+}(\pi^0 n)\end{Bmatrix}\label{17}.
\end{align}
The experimental value for $E_{0+}(\pi^-p)$ taken from \cite{KO97} and the leading term of its chiral expansion 
calculated in \cite{BKM96a} up to $\Order\big(e p^3\big)$ are
\beq
E_{0+}(\pi^- p)=-\frac{\sqrt{2}\,e\ga}{8\pi\Fpi}+\Order(e p)=(-31.5 \pm 0.8)\cdot 10^{-3}\mpi^{-1},
\eeq
whereas $E_{0+}(\pi^0n)$ only starts at $\Order\big(e p^2\big)$ (explicit expressions are given in \cite{BKM94});
therefore both contributions to $\text{Im}\,a_{\pi^- p}^{\rm cex}$ in~\eqref{17} 
are suppressed by at least one chiral order.
Unfortunately, $E_{0+}(\pi^0n)$ is not directly accessible in experiment. 
Combining deuterium data with ChPT predictions \cite{BBLM97} yields
\beq
E_{0+}(\pi^0 n)=(2.1\pm 0.5)\cdot 10^{-3}\mpi^{-1}.
\eeq
Inserting the chiral expansions of $E_{0+}(\pi^-p)$ and $a^-$ into \eqref{17} reproduces the imaginary part of $a_{\pi^- p}$ 
appearing in \eqref{eq:Delta_a}. 
One can easily check that the difference between \eqref{17} and its chiral expansion is mainly due to $\Delta_{\rm N}$, 
which justifies our treatment of the cusp effect in Sect.~\ref{sec:analytic}.
Separating strong (first number) and electromagnetic contributions, we obtain numerically
\begin{align}
\text{Im}\,a_{\pi^- p}&= \Big((2.91 \pm 0.12) + (1.86 \pm 0.09) \Big)\cdot 10^{-3}\mpi^{-1}\notag\\
&=(4.77 \pm 0.15)\cdot 10^{-3}\mpi^{-1},\\
\text{Im}\,a_{\pi^- p}^{\rm cex}&= \Big((-0.04 \pm 0.05) +(-0.12 \pm 0.03)\Big)\cdot 10^{-3}\mpi^{-1}\notag\\
&=(-0.16 \pm 0.06)\cdot 10^{-3}\mpi^{-1} \notag.
\end{align}

\begin{sloppypar}
Finally, the above results may be checked based on the observation that the ratio
\beq
\frac{\left.\text{Im}\,a_{\pi^- p}\right|_{\rm strong}}{\left.\text{Im}\,a_{\pi^- p}\right|_{\rm EM}}=1.57 \pm 0.10
\eeq
should correspond to the so-called Panofsky ratio
$
P={\sigma(\pi^-p\rightarrow \pi^0 n)}/{\sigma(\pi^-p\rightarrow \gamma n)}.
$
Indeed, its experimental value is found to be $P=1.546\pm 0.009$ (cf.~\cite{GLR07,Spuller:1977ve}).
\end{sloppypar}

\section{Comparison to earlier work}
\label{sec:comp}

Our result for $\pi^-p\rightarrow \pi^- p$ agrees with \cite{GR02}. 
In~\cite{FMS99}, a similar analysis of isospin breaking was performed in heavy-baryon ChPT, 
switching off virtual photons. This corresponds to 
\beq
\Delta_\pi \neq 0, \ \Delta_{\rm N} \neq 0, \ f_1 e^2\neq 0, \ f_2 e^2\neq 0, \ \md \neq \muu,\  e^2=0.
\eeq
Furthermore, isospin-breaking effects due to the Dirac spinors are neglected and 
$B_{\rm thr}^-$ is expressed by $c_4$. 
We have checked explicitly for the triangle relation, 
for the charge exchange reactions, and for the neutral-pion elastic channels, that our results coincide in this limit.
Numerically, we find $R = (0.74\pm 0.21)\,\%$, which is compatible with the numerical value 
$R_{\rm FMS}=(0.9\dots 1.1)\,\%$ quoted in \cite{FMS99}. 
Both values slightly differ, since the denominator is not expressed by $a^-$ 
(the additional LECs needed are taken from \cite{FMS98}) 
and since the isospin limit is defined as the average between charged and neutral particles.

Virtual photons were taken into account in \cite{FM01} in order to study isospin violation above threshold. 
Unfortunately, a direct comparison is not possible, as no analytic expressions for the amplitudes are provided. 
Even more, also a numerical comparison is difficult due to a conceptual difference: 
in \cite{FM01}, the electromagnetic corrections were used to pin down the LECs from experimental data, 
and thereafter applied to extract the strong amplitude. 
In particular, electromagnetic contributions to the particle masses were switched off. 
Thus, the quoted isospin-breaking effect of $-0.7\,\%$ for the triangle relation in the S-wave 
refers to strong isospin violation only.

\section{Summary and outlook}

In this letter, we have systematically analyzed isospin violation in the $\pi N$ scattering
lengths in all channels, including a detailed estimate of the
theoretical uncertainties. 
The extension of this analysis beyond threshold will be the subject of future work,
to which we also refer for details of the calculation~\cite{abovethresh}.

We find that isospin violation is quite small in $\pi^- p\rightarrow \pi^0 n$, 
at the order of one percent at most, whereas the
charged-pion elastic channels display more sizeable effects on the few-percent level. 
In particular, the so-called triangle relation that vanishes in the isospin limit
is violated by about 1.5\% consistent with earlier findings in heavy-baryon
ChPT and inconsistent with the 5--7\% deviation extracted from the data at
lowest pion momenta in \cite{Gibbs:1995dm,Matsinos:1997pb}.
In addition, we find a substantial isospin-breaking correction to the neutral-pion--proton scattering
length. In view of these results, further experiments e.g.\ at HI$\gamma$S and MAMI are urgently called for.

\section*{Acknowledgements}
Partial financial support by the Helmholtz Association through funds provided
to the virtual institute ``Spin and strong QCD'' (VH-VI-231),
by the European Community-Research Infrastructure Integrating Activity 
``Study of Strongly Interacting Matter''
(acronym HadronPhysics2, Grant Agreement n.~227431) under the Seventh 
Framework Programme of the EU,
and by DFG (SFB/TR 16, ``Subnuclear Structure of Matter'') is gratefully
acknowledged. 

\appendix

\section{Effective Lagrangians}
\label{app:lagr}

We will use the effective Lagrangian for nucleons, pions, and virtual photons, as constructed in~\cite{GR02}, 
whereof we actually need the following terms:
\begin{align}
\Lagr_{\rm eff}&=\sum_{i=1}^2\Bigl(\Lagr_\pi^{(p^{2i})}+\Lagr_\pi^{(e^2p^{2i-2})}\Bigr)
+\sum_{i=1}^3\Lagr_{\rm N}^{(p^i)}+\sum_{i=0}^1\Lagr_{\rm N}^{(e^2p^i)}+\Lagr_\gamma,\notag\\
\Lagr_\pi^{(p^2)}&+\Lagr_\pi^{(e^2)}+\Lagr_\gamma=\frac{F^2}{4}\langle d^\mu U^\dagger d_\mu U+\chi^\dagger U+U^\dagger \chi\rangle\notag\\
&+Z F^4\langle \mathcal{Q}U\mathcal{Q}U^\dagger\rangle-\frac{1}{4}F_{\mu\nu}F^{\mu\nu}-\frac{1}{2}\big(\partial_\mu A^\mu\big)^2,\notag\\
\Lagr_\pi^{(p^4)}&=\frac{l_4}{4}\langle d^\mu U^\dagger d_\mu \chi+d^\mu \chi^\dagger d_\mu U \rangle ,\notag\\
\Lagr_\pi^{(e^2p^2)}&=F^2 \Bigl\{\langle d^\mu U^\dagger d_\mu U \rangle \bigl(k_1\langle \mathcal{Q}^2 \rangle
+k_2 \langle \mathcal{Q}U\mathcal{Q}U^\dagger \rangle\bigr)\notag\\
&+k_3\bigl(\langle d^\mu U^\dagger \mathcal{Q} U \rangle\langle d_\mu U^\dagger \mathcal{Q} U \rangle+
 \langle d^\mu U \mathcal{Q} U^\dagger\rangle\langle d_\mu U \mathcal{Q} U^\dagger\rangle\bigr)
\notag\\
&+k_4\langle d^\mu U^\dagger \mathcal{Q} U \rangle \langle d_\mu U \mathcal{Q} U^\dagger \rangle\Bigr\} ,\notag\\
\Lagr_{\rm N}^{(p)}&=\bar{\Psi}\Big\{i\slashed{D}-m+\frac{1}{2}g \slashed{u}\gamma_5\Big\}\Psi, \notag\\
\Lagr_{\rm N}^{(p^2)}&=\bar{\Psi}\Big\{c_1 \langle\chi_+\rangle -\frac{c_2}{4m^2}\langle u_\mu u_\nu\rangle D^\mu D^\nu + {\rm h.c.}\notag\\
&+\frac{c_3}{2}\langle u_\mu u^\mu\rangle+\frac{i}{4}c_4 \sigma^{\mu\nu}[u_\mu,u_\nu]+c_5 \hat{\chi}_+\Big\}\Psi\notag,\\
\Lagr_{\rm N}^{(e^2)}&=F^2\bar{\Psi}\Big\{f_{1/3} \langle \hat{Q}^2_+ \mp Q_-^2\rangle+ f_2\langle Q_+\rangle \hat{Q}_+\Big\}\Psi,\notag\\
\Lagr_{\rm N}^{(p^3)}&=\bar{\Psi}\Big\{-\frac{d_1}{2m}[u_\mu,[D_\nu,u^\mu]]D^\nu
-\frac{d_2}{2m}[u_\mu,[D^\mu,u_\nu]]D^\nu\notag\\
&+\frac{d_3}{12m^3}[u_\mu,[D_\nu,u_\lambda]]\bigl(D^\mu D^\nu D^\lambda+{\rm sym}\bigr)\notag\\
&+\frac{i}{2m}d_5[\chi_-,u_\mu]D^\mu\Big\}\Psi+{\rm h.c.},\\
\Lagr_{\rm N}^{(e^2p)}&=\frac{iF^2}{2m}\bar{\Psi}\Big\{g_6\langle Q_+\rangle \langle Q_-u_\mu\rangle D^\mu 
+g_{7/8} \langle Q_\pm u_\mu\rangle Q_\mp D^\mu\Big\}\Psi+{\rm h.c.},\notag
\end{align}
where $\langle A\rangle$ denotes the trace of a matrix $A$, $\hat{A}=A-\langle A\rangle/2$ its traceless part,
$
\bar{\Psi}(\mathcal{O}+{\rm h.c.})\Psi\equiv \bar{\Psi}\mathcal{O}\Psi+{\rm h.c.}
$
for an operator $\mathcal{O}$ and
\begin{align}
 d_\mu U&=\partial_\mu U-i A_\mu[\mathcal{Q},U], \quad \chi=2 B\,\text{diag}(\muu,\md), \quad U=u^2,\notag\\
F_{\mu\nu}&=\partial_\mu A_\nu-\partial_\nu A_\mu, \quad \mathcal{Q}=\frac{e}{3}\,\text{diag}(2,-1),\quad  Q=e\,\text{diag}(1,0),\notag\\
D_\mu &=\partial_\mu + \Gamma_\mu, \quad  \Gamma_\mu= \frac{1}{2}\Big(u^\dagger(\partial_\mu-i Q A_\mu)u+u(\partial_\mu-i Q A_\mu)u^\dagger\Big),\notag\\
\chi_\pm&=u^\dagger \chi u^\dagger\pm u \chi^\dagger u,\  u_\mu= i\Big(u^\dagger(\partial_\mu\!-\!i Q A_\mu)u-u(\partial_\mu\!-\!i Q A_\mu)u^\dagger\Big),\notag\\
Q_\pm&=\frac{1}{2}(u Q u^\dagger \pm u^\dagger Q u), \quad [D_\mu,u_\nu]=\partial_\mu u_\nu+[\Gamma_\mu,u_\nu].
\end{align}
$\Psi=(p,n)^T$ contains the nucleon fields and the matrix $U$ 
collects the pion fields in the usual way.
$F$ is the pion decay constant in the chiral limit and is replaced by its physical value $\Fpi$ by means of
\beq
\Fpi=F\left\{1+\frac{\mpi^2}{F^2}\left(l_4^{\rm r}-\frac{1}{16\pi^2}\log \frac{\mpi^2}{\mu^2}\right)\right\}+\Order\big(\mpi^4\big),
\eeq
while the chiral-limit axial charge $g$ may be identified with its physical value $\ga$, 
since axial contributions only start at $\Order(p^2)$ at threshold. $m$ denotes the nucleon mass in the chiral limit.

Changing the version of $\Lagr_\pi^{(p^4)}$ from \cite{GL84} to \cite{GSS88} results in a redefinition of $d_5^{\rm r}$. 
This $\tilde{d}_5^{\rm r}$ is related to our $d_5^{\rm r}$ by  
\beq
\Fpi^2\tilde{d}_5^{\rm r}(\mu)=\Fpi^2d_5^{\rm r}(\mu)+\frac{1}{8}l_4^{\rm r}(\mu).
\eeq
Note that in this convention $l_4^{\rm r}$ disappears in \eqref{1}.

\section{Pion--neutron scattering lengths}
\label{app:chann}

The strong contributions to the remaining channels are determined by charge symmetry
(the discrete subgroup of the general isospin transformations that only exchanges
$u\leftrightarrow d$ on the quark level), such that 
only the electromagnetic parts have to be calculated explicitly: the pion mass difference alone
cannot contribute to charge-symmetry breaking. 
[How to simplify a calculation of isospin-breaking 
effects by such considerations is explained in more detail in~\cite{KL06}.]
The results are 
\begin{align}
\Delta a_{\pi^+ n}&=a_{\pi^+ n}-(a^++a^-)=\Big(-4.3^{+4.3}_{-6.5}+6.0 i\Big)\cdot 10^{-3}\mpi^{-1}\notag\\
&= \Delta a_{\pi^- p} + \frac{e^2\mpp}{4\pi(\mpp+\mpi)} \Biggl\{f_2 - 2\mpi\left(2g^{\rm r}_6+g^{\rm r}_8\right)\notag\\
&+i\frac{\mpi^2}{8\pi\Fpi^4}\Big(\sqrt{\Delta_\pi+2\mpi\Delta_{\rm N}}-\sqrt{\Delta_\pi-2\mpi\Delta_{\rm N}}\Big)\Biggr\},\notag\\
\Delta a_{\pi^- n}&=a_{\pi^- n}-(a^+-a^-)=-6.2^{+4.3}_{-6.5}\cdot 10^{-3}\mpi^{-1} \notag\\
&=\Delta a_{\pi^+ p} + \frac{e^2\mpp}{4\pi(\mpp+\mpi)} \biggl\{f_2 + 2\mpi\left(2g^{\rm r}_6+g^{\rm r}_8\right)\biggr\},\notag\\
\Delta a_{\pi^+ n}^{\rm cex}&=a_{\pi^+ n}^{\rm cex}+\sqrt{2}\,a^-=(2.3 \pm 0.9)\cdot 10^{-3}\mpi^{-1} \notag\\
&=\Delta a_{\pi^- p}^{\rm cex}+\frac{\sqrt{2}\,\mpp}{4\pi(\mpp+\mpi)}\Biggl\{\frac{\mpi\Delta_{\rm N}}{2\Fpi^2\mpp}\big(1+2\ga^2\big)-e^2 f_2\Biggr\},\notag\\
\Delta a_{\pi^0 n}&=a_{\pi^0 n}-a^+=(-1.8 \pm 0.2)\cdot 10^{-3}\mpi^{-1}\notag\\
&=\Delta a_{\pi^0 p} + \frac{\mpp}{4\pi(\mpp+\mpi)} \Biggl\{-\frac{4c_5 B(\md-\muu)}{\Fpi^2}\notag\\
&+\frac{\mpi^2}{8\pi\Fpi^4}\Big(\sqrt{\Delta_\pi+2\mpi\Delta_{\rm N}}-\sqrt{\Delta_\pi-2\mpi\Delta_{\rm N}}\Big)\Biggr\}.
\end{align}
$a_{\pi^+ n}^{\rm cex}$, which is accessible through the cusp in neutral-pion photoproduction on the proton,
receives only moderate isospin-breaking corrections ($(-1.9\pm0.8)\,\%$), whose uncertainty
is rather well-controlled.
The correction to the two-body contribution to $\text{Re}\,a_{\pi d} \propto 2(a^+ + \Delta \tilde a^+)+\ldots$
displays the same dependence on $f_1$ and $c_1$ as $2(a^++\Delta a^+)$. It is determined by
\begin{align}
\Delta \tilde a^+ &= \frac{\mpp}{4\pi(\mpp+\mpi)}
\Bigg\{\!\frac{4\Delta_\pi}{\Fpi^2}c_1-e^2\Big(2f_1 - \mpi\big(2g^{\rm r}_6+g^{\rm r}_8\big)\Big) \notag\\
&-\frac{\ga^2\mpi}{32\pi\Fpi^2}\biggl(\frac{33\Delta_\pi}{4\Fpi^2} +e^2\biggr) \Bigg\} ,
\end{align} 
which, in addition to $f_1$ and $c_1$, also includes a sizeable fixed shift (cf.\ the discussion in~\cite{BH07,GLR09}).
Finally,
we point out that the remnants of the cusp effect contribute roughly one third to the difference
$a_{\pi^0 p}-a_{\pi^0 n}=(-3.4\pm0.4)\cdot 10^{-3}\mpi^{-1}$ and hence modify Weinberg's prediction~\cite{Weinberg77}
significantly.  
[This is in apparent contrast to the finding in~\cite{MM99} where the complete $\Order(p^4)$
corrections to $a_{\pi^0 p}-a_{\pi^0 n}$ have been calculated; 
however, the result for the cusp is incorrect.]


\begin{thebibliography}{99}
\biboptions{sort&compress}

\bibitem{Weinberg77}
  S.~Weinberg,
  %``The Problem Of Mass,''
  Trans.\ New York Acad.\ Sci.\  {\bf 38} (1977) 185.
  %%CITATION = TNYAA,38,185;%%

\bibitem{MS97}
  U.-G.~Mei{\ss}ner and S.~Steininger,
  %``Isospin violation in pion-nucleon scattering,''
  Phys.\ Lett.\  B {\bf 419} (1998) 403
  [arXiv:hep-ph/9709453].
  %%CITATION = PHLTA,B419,403;%%

\bibitem{FMS99}
  N.~Fettes, U.-G.~Mei{\ss}ner and S.~Steininger,
  %``On the size of isospin violation in low-energy pion nucleon scattering,''
  Phys.\ Lett.\  B {\bf 451} (1999) 233
  [arXiv:hep-ph/9811366].
  %%CITATION = PHLTA,B451,233;%%

\bibitem{MM99}
  G.~M\"uller and U.-G.~Mei{\ss}ner,
  %``Virtual photons in baryon chiral perturbation theory,''
  Nucl.\ Phys.\  B {\bf 556} (1999) 265
  [arXiv:hep-ph/9903375].
  %%CITATION = NUPHA,B556,265;%%

\bibitem{FM01b}
  N.~Fettes and U.-G.~Mei{\ss}ner,
  %``Towards an understanding of isospin violation in pion nucleon
  %scattering,''
  Phys.\ Rev.\  C {\bf 63} (2001) 045201
  [arXiv:hep-ph/0008181].
  %%CITATION = PHRVA,C63,045201;%%
    
\bibitem{FM01}
  N.~Fettes and U.-G.~Mei{\ss}ner,
  %``Complete analysis of pion nucleon scattering in chiral perturbation  theory
  %to third order,''
  Nucl.\ Phys.\  A {\bf 693} (2001) 693
  [arXiv:hep-ph/0101030].
  %%CITATION = NUPHA,A693,693;%%

\bibitem{GR02}
  J.~Gasser, M.~A.~Ivanov, E.~Lipartia, M.~Moj\v{z}i\v{s} and A.~Rusetsky,
  %``Ground-state energy of pionic hydrogen to one loop,''
  Eur.\ Phys.\ J.\  C {\bf 26} (2002) 13
  [arXiv:hep-ph/0206068].
  %%CITATION = EPHJA,C26,13;%%
  
\bibitem{BL99}
  T.~Becher and H.~Leutwyler,
  %``Baryon chiral perturbation theory in manifestly Lorentz invariant form,''
  Eur.\ Phys.\ J.\  C {\bf 9} (1999) 643
  [arXiv:hep-ph/9901384].
  %%CITATION = EPHJA,C9,643;%%

\bibitem{Bernard:2007zu}
  V.~Bernard,
  %``Chiral Perturbation Theory and Baryon Properties,''
  Prog.\ Part.\ Nucl.\ Phys.\  {\bf 60} (2008) 82
  [arXiv:0706.0312 [hep-ph]].
  %%CITATION = PPNPD,60,82;%%

\bibitem{MRR06}
  U.-G.~Mei{\ss}ner, U.~Raha and A.~Rusetsky,
  %``Isospin-breaking corrections in the pion deuteron scattering length,''
  Phys.\ Lett.\  B {\bf 639} (2006) 478
  [arXiv:nucl-th/0512035].
  %%CITATION = PHLTA,B639,478;%%

\bibitem{Bernstein:1998ip}
  A.~M.~Bernstein,
  %``Light quark mass difference and isospin breaking in electromagnetic  pion
  %production,''
  Phys.\ Lett.\  B {\bf 442} (1998) 20
  [arXiv:hep-ph/9810376].
  %%CITATION = PHLTA,B442,20;%%

\bibitem{Bernstein:2009dc}
  A.~M.~Bernstein, M.~W.~Ahmed, S.~Stave, Y.~K.~Wu and H.~R.~Weller,
  %``Chiral Dynamics in Photo-Pion Physics: Theory, Experiment, and Future
  %Studies at the HI$\gamma$S Facility,''
  arXiv:0902.3650 [nucl-ex].
  %%CITATION = ARXIV:0902.3650;%%
 
\bibitem{BL01}
  T.~Becher and H.~Leutwyler,
  %``Low energy analysis of pi N --> pi N,''
  JHEP {\bf 0106} (2001) 017
  [arXiv:hep-ph/0103263].
  %%CITATION = JHEPA,0106,017;%% 
  
\bibitem{BKM93}
  V.~Bernard, N.~Kaiser and U.-G.~Mei{\ss}ner,
  %``Chiral corrections to the S wave pion - nucleon scattering lengths,''
  Phys.\ Lett.\  B {\bf 309} (1993) 421
  [arXiv:hep-ph/9304275].
  %%CITATION = PHLTA,B309,421;%%
  
\bibitem{GLR01}
  J.~Gasser, V.~E.~Lyubovitskij, A.~Rusetsky and A.~Gall,
  %``Decays of the pi+ pi- atom,''
  Phys.\ Rev.\  D {\bf 64} (2001) 016008
  [arXiv:hep-ph/0103157].
  %%CITATION = PHRVA,D64,016008;%%

\bibitem{BFGKB09}
  M.~Bissegger, A.~Fuhrer, J.~Gasser, B.~Kubis and A.~Rusetsky,
  %``Radiative corrections in K --> 3 pi decays,''
  Nucl.\ Phys.\  B {\bf 806} (2009) 178
  [arXiv:0807.0515 [hep-ph]].
  %%CITATION = NUPHA,B806,178;%%

\bibitem{GL82}
  J.~Gasser and H.~Leutwyler,
  %``Quark Masses,''
  Phys.\ Rept.\  {\bf 87} (1982) 77.
  %%CITATION = PRPLC,87,77;%%

\bibitem{M05}
  U.-G.~Mei{\ss}ner,
  %``Quark mass dependence of baryon properties,''
  PoS {\bf LAT2005} (2006) 009
  [arXiv:hep-lat/0509029].
  %%CITATION = POSCI,LAT2005,009;%%

\bibitem{MOJ98}
  M.~Moj\v{z}i\v{s},
  %``Elastic pi N scattering to O(p**3) in heavy baryon chiral perturbation
  %theory,''
  Eur.\ Phys.\ J.\  C {\bf 2} (1998) 181
  [arXiv:hep-ph/9704415].
  %%CITATION = EPHJA,C2,181;%%

\bibitem{BM99}
  P.~B{\"u}ttiker and U.-G.~Mei{\ss}ner,
  %``Pion nucleon scattering inside the Mandelstam triangle,''
  Nucl.\ Phys.\  A {\bf 668} (2000) 97
  [arXiv:hep-ph/9908247].
  %%CITATION = NUPHA,A668,97;%%

\vfill\eject

\bibitem{FMS98}
  N.~Fettes, U.-G.~Mei{\ss}ner and S.~Steininger,
  %``Pion nucleon scattering in chiral perturbation theory.  I:
  %Isospin-symmetric case,''
  Nucl.\ Phys.\  A {\bf 640} (1998) 199
  [arXiv:hep-ph/9803266].
  %%CITATION = NUPHA,A640,199;%%

\bibitem{F00}
  N.~Fettes, 
  %``Pion nucleon physics in chiral perturbation theory,''
  PhD thesis, Berichte des FZ J\"ulich, J\"ul-3814.
  %%CITATION = JUEL-3814;%%
  
\bibitem{FM00}
  N.~Fettes and U.-G.~Mei{\ss}ner,
  %``Pion nucleon scattering in chiral perturbation theory. II: Fourth order
  %calculation,''
  Nucl.\ Phys.\  A {\bf 676} (2000) 311
  [arXiv:hep-ph/0002162].
  %%CITATION = NUPHA,A676,311;%%
  
\bibitem{Ko80}
  R.~Koch and E.~Pietarinen,
  %``Low-Energy Pi N Partial Wave Analysis,''
  Nucl.\ Phys.\  A {\bf 336} (1980) 331.
  %%CITATION = NUPHA,A336,331;%%

\bibitem{HIS07}
  C.~Haefeli, M.~A.~Ivanov and M.~Schmid,
  %``Electromagnetic low-energy constants in ChPT,''
  Eur.\ Phys.\ J.\  C {\bf 53} (2008) 549
  [arXiv:0710.5432 [hep-ph]].
  %%CITATION = EPHJA,C53,549;%%
  
\bibitem{MOU97}
  B.~Moussallam,
  %``A sum rule approach to the violation of Dashen's theorem,''
  Nucl.\ Phys.\  B {\bf 504} (1997) 381
  [arXiv:hep-ph/9701400].
  %%CITATION = NUPHA,B504,381;%%

\bibitem{AMOU04}
  B.~Ananthanarayan and B.~Moussallam,
  %``Four-point correlator constraints on electromagnetic chiral parameters  and
  %resonance effective Lagrangians,''
  JHEP {\bf 0406} (2004) 047
  [arXiv:hep-ph/0405206].
  %%CITATION = JHEPA,0406,047;%%

\bibitem{PDG08}
  C.~Amsler {\it et al.}  [Particle Data Group],
  %``Review of particle physics,''
  Phys.\ Lett.\  B {\bf 667} (2008) 1.
  %%CITATION = PHLTA,B667,1;%%
 
\bibitem{KO97}
  M.~A.~Kovash  [E643 Collaboration],
  %``Total cross-sections for pi- p $\to$ gamma n at 10-MeV to 20-MeV,''
  PiN Newslett.\  {\bf 12N3} (1997) 51.
  %%CITATION = 00076,12N3,51;%%
  
\bibitem{BKM96a}
  V.~Bernard, N.~Kaiser and U.-G.~Mei{\ss}ner,
  %``Chiral corrections to the Kroll-Ruderman theorem,''
  Phys.\ Lett.\  B {\bf 383} (1996) 116
  [arXiv:hep-ph/9603278].
  %%CITATION = PHLTA,B383,116;%%

\bibitem{BKM94}
  V.~Bernard, N.~Kaiser and U.-G.~Mei{\ss}ner,
  %``Neutral pion photoproduction off nucleons revisited,''
  Z.\ Phys.\  C {\bf 70} (1996) 483
  [arXiv:hep-ph/9411287].
  %%CITATION = ZEPYA,C70,483;%%

\bibitem{BBLM97}
  S.~R.~Beane, V.~Bernard, T.~S.~H.~Lee, U.-G.~Mei{\ss}ner and U.~van Kolck,
  %``Neutral pion photoproduction on deuterium in baryon chiral perturbation
  %theory to order q**4,''
  Nucl.\ Phys.\  A {\bf 618} (1997) 381
  [arXiv:hep-ph/9702226].
  %%CITATION = NUPHA,A618,381;%%
  
\bibitem{GLR07}
  J.~Gasser, V.~E.~Lyubovitskij and A.~Rusetsky,
  %``Hadronic atoms in QCD + QED,''
  Phys.\ Rept.\  {\bf 456} (2008) 167
  [arXiv:0711.3522 [hep-ph]].
  %%CITATION = PRPLC,456,167;%%

\bibitem{Spuller:1977ve}
  J.~Spuller {\it et al.},
  %``A Remeasurement Of The Panofsky Ratio,''
  Phys.\ Lett.\  B {\bf 67} (1977) 479.
  %%CITATION = PHLTA,B67,479;%%

\bibitem{abovethresh}
  M.~Hoferichter, B.~Kubis and U.-G.~Mei{\ss}ner, in preparation.

\bibitem{Gibbs:1995dm}
  W.~R.~Gibbs, L.~Ai and W.~B.~Kaufmann,
  %``Isospin Breaking In Low-Energy Pion Nucleon Scattering,''
  Phys.\ Rev.\ Lett.\  {\bf 74} (1995) 3740.
  %%CITATION = PRLTA,74,3740;%%

\bibitem{Matsinos:1997pb}
  E.~Matsinos,
  %``Isospin violation in the pi N system at low energies,''
  Phys.\ Rev.\  C {\bf 56} (1997) 3014.
  %%CITATION = PHRVA,C56,3014;%%

\bibitem{GL84}
  J.~Gasser and H.~Leutwyler,
  %``Chiral Perturbation Theory To One Loop,''
  Annals Phys.\  {\bf 158} (1984) 142.
  %%CITATION = APNYA,158,142;%%

\bibitem{GSS88}
  J.~Gasser, M.~E.~Sainio and A.~\v{S}varc,
  %``Nucleons With Chiral Loops,''
  Nucl.\ Phys.\  B {\bf 307} (1988) 779.
  %%CITATION = NUPHA,B307,779;%%

\bibitem{KL06}
  B.~Kubis and R.~Lewis,
  %``Isospin violation in the vector form factors of the nucleon,''
  Phys.\ Rev.\  C {\bf 74} (2006) 015204
  [arXiv:nucl-th/0605006].
  %%CITATION = PHRVA,C74,015204;%%

\bibitem{BH07}
  V.~Baru, J.~Haidenbauer, C.~Hanhart, A.~E.~Kudryavtsev, V.~Lensky and U.-G.~Mei{\ss}ner,
  %``Pion--deuteron scattering length in Chiral Perturbation Theory up to order
  %\chi^{3/2},''
  {\it Proc.\ 11th Int.\ Conf.\ on Meson-Nucleon Physics and the Structure of the Nucleon (MENU 2007), J\"ulich}
  [arXiv:0711.2743 [nucl-th]].
  %%CITATION = ECONF,C070910,127;%%

\bibitem{GLR09}
  J.~Gasser, V.~E.~Lyubovitskij and A.~Rusetsky,
  %``Hadronic Atoms,''
  arXiv:0903.0257 [hep-ph].
  %%CITATION = ARXIV:0903.0257;%%

\end{thebibliography}
\end{document}